\documentclass[twocolumn,aps,prb,superscriptaddress,amssymb,showpacs]{revtex4}

\usepackage{graphicx}
\usepackage{dcolumn}
\usepackage{amsmath}

\begin{document}

\title{Asymmetric Reflectance and Cluster Size Effects in Silver Percolation Films}

\author{Nicholas A. Kuhta}
\email{kuhtan@onid.orst.edu}
\affiliation{Department of Physics, 301 Weniger Hall, Oregon State
University, Corvallis, OR 97331, USA}

\author{Aiqing Chen}
\altaffiliation[Present address: ]{Material Science Division, Argonne National Laboratory, Argonne, IL 60439, USA}
\affiliation{Department of Physics, 1274 University of Oregon, Eugene, Oregon 97403, USA}

\author{Keisuke Hasegawa}
\altaffiliation[Present address: ]{Laboratory of Cellular and Molecular Biology, National Cancer Institute, NIH, Bethesda, MD 20892-4256, USA}
\affiliation{Department of Physics, 1274 University of Oregon, Eugene, Oregon 97403, USA}

\author{Miriam Deutsch}
\email{miriamd@uoregon.edu} \affiliation{Department of Physics, 1274 University of Oregon, Eugene, Oregon 97403, USA}

\author{Viktor A. Podolskiy}
\email{viktor_podolskiy@uml.edu}
\affiliation{Department of Physics and Applied Physics, One University Avenue, University of Massachusetts Lowell, Lowell, MA 01854, USA}
\affiliation{Department of Physics, 301 Weniger Hall, Oregon State
University, Corvallis, OR 97331, USA}

\begin{abstract}
We develop a quantitative description of giant asymmetry in reflectance, recently observed in semicontinuous metal films. The developed scaling-theory based technique reproduces the spectral properties of semicontinuous composites, as well as provides insight into the origin of experimentally observed loss, reflectance, and transmittance anomalies in the vicinity of the percolation threshold.
\end{abstract}

\pacs{42.25.Dd,78.67.Sc,78.67.Bf,78.66.Sq}

\maketitle

\section{Introduction}
\begin{figure}[t]
\includegraphics[width=5.45cm]{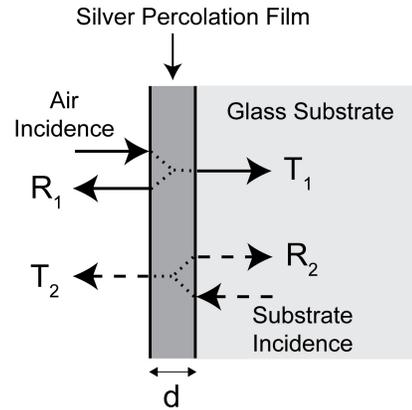}
\caption{(Color online) General layered structure composed of a silver percolation film clad by air to the left and glass to the right.  Incident light may come from either the air or substrate side as shown.  Both air and glass regions are taken to be semi-infinite.
 \label{FilmFigure} }
\end{figure}

Research into the optics of semicontinuous metal-dielectric films has been enjoying sustained interest due to a unique combination of novel physics and the practical applications offered by such composites.  It has been demonstrated, both theoretically and experimentally, that the electromagnetic (EM) response of these structures is dominated by a non-trivial interplay between Anderson-localized and delocalized surface plasmon polaritons.\cite{StockmanLocal,genov,akslocal} This results in unusual optical properties that include: greatly enhanced absorption, giant intensity fluctuations of local EM fields, giant local chiral response, and strongly enhanced optical nonlinearities.\cite{physreports,shalaevNL,optactiv,lagarkov,shalaevbook,Stockmanprl2000,Stockmanprl1994} In a related context, it has been recently shown that the reflectance of semicontinuous silver nanocomposites, chemically deposited on glass substrates, strongly depends on the direction of incident light.\cite{chenMM}  In particular, the reflectance of such a system irradiated from the substrate/film interface side can differ by as much as $15\%$ from its reflectance given film/air side incidence (see Fig.\ref{FilmFigure}). Moreover, this large asymmetry in reflectance has been found to be extremely broadband, spanning most of the visible frequency spectrum. For comparison, the reflectance asymmetry of thin, \emph{continuous} silver films does not exceed $3\%$ when measured over the same range of optical frequencies, and does not exhibit any broadband characteristics. It has been suggested that the origin of this large broadband asymmetry is in the enhanced optical absorbance which is often seen in percolation-type systems. Here we develop a quantitative description of the observed phenomenon.

The geometry of the system described in this manuscript is shown in Figure \ref{FilmFigure}.  We approximate the silver percolation film as a uniform material with thickness $d$ [in our calculations $d=50$nm (see section III)]. The microstructure of the film is characterized by the surface metal filling fraction $p$ ranging from $p=0$ for a bare glass substrate to $p=1$ for a substrate which is fully covered with metal.  At the critical value $p=p_c$, known as the \emph{percolation threshold}, the $dc$ conductivity response of the entire random metal-dielectric composite undergoes an insulator-conductor phase transition.\cite{intropt} The unique optical properties of our films are manifest particularly in the vicinity of the percolation threshold, and we therefore adopt the conventional description of the response as function of the parameter $p-p_c$ for the purpose of both modeling and data analysis.  All theoretical and experimental results in this work use $p_c=0.6$, which correlates well with two dimensional site percolation on a square lattice ($p_c\simeq0.593$).\cite{intropt}

As mentioned above, the reflectance $R_1$ of the composite film, measured using light impinging from the air/metal interface, strongly differs from $R_2$ - the reflectance measured when light is incident from the substrate/film side.  Since the transmittance of our system, as the transmittance of any non-chiral homogeneous film is symmetric (\emph{i.e.} $T_1=T_2$)\cite{ChiralBook,chiralprl}, the asymmetry in reflectance $\Delta R\equiv R_1-R_2$ directly reflects the asymmetry in losses.\cite{footnoteLoss}  As we show below, in contrast to vacuum-deposited percolation films, (i) $\Delta R$ as well as the computed combined losses exhibit a local minimum at $p\simeq p_c$, (ii) $\Delta R$ exhibits broadband response in the vicinity of $p-p_c\simeq\pm0.05$ , (iii) the reflectance exhibits a local maximum in the vicinity of $p\simeq p_c$, and (iv) the transmittance exhibits a local minimum near $p\simeq p_c$.

\section{percolation film synthesis and characterization}
\begin{figure}[h]
\includegraphics[width=8.0cm]{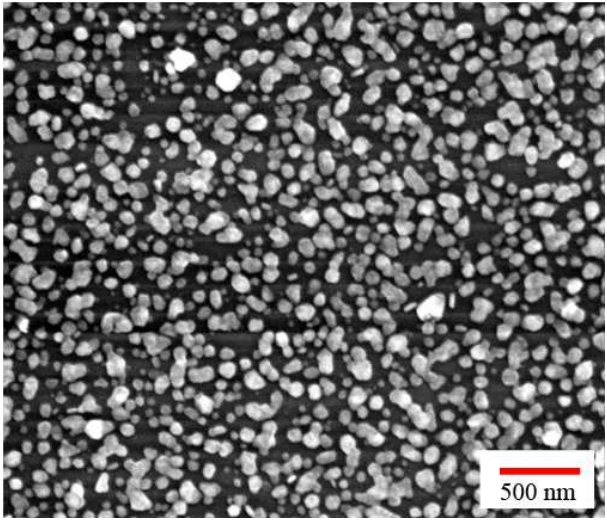}
\caption{(Color online) Scanning electron micrograph of a chemically deposited silver film with metal filling fraction $p\simeq0.52$. The scale bar is 500nm.
 \label{SEMfig} }
\end{figure}

Semi-continuous silver films with controllable filling fractions were deposited on microscope slides using a modified Tollen's reaction as described previously.\cite{chenMM,JAPfilms} The amount of silver deposited on the substrates was controlled by monitoring deposition times, with reactions ranging between 1-6 h. Ensuing deposition, coated substrates were rinsed with ultrapure water before being dried with filtered air and stored under nitrogen until tested. Figure \ref{SEMfig} shows a scanning electron micrograph of a typical film with filling fraction $p\simeq0.52$. These chemically deposited films appear as highly disordered polycrystalline aggregates, with large grain size distributions. In addition, we note the non-uniform coating of the substrates by the metal, resulting in  highly discontinuous morphologies.

\begin{figure}[h]
\includegraphics[width=\columnwidth]{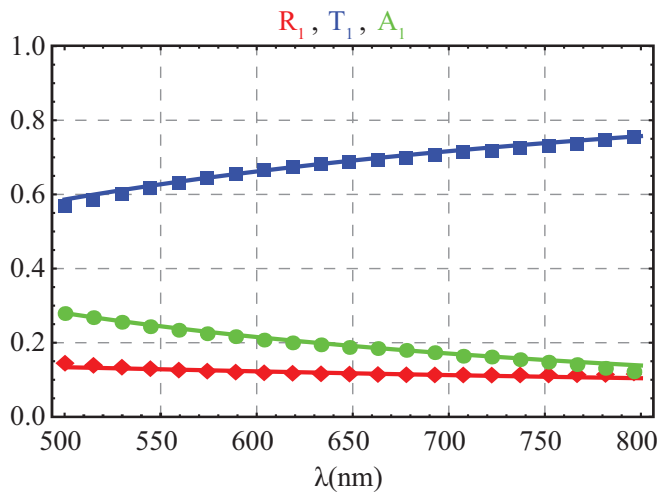}
\caption{(Color online) Measured reflectance (red diamonds), transmittance (blue boxes), and absorbance (green circles) as function of incident wavelength for measured metal filling fraction $p\simeq0.52$.  Solid lines represent the results of scaling theory calculations.
 \label{RTAfig} }
\end{figure}

Optical reflectance and transmittance spectra were collected using a spectroscopic optical microscopy setup.\cite{chenMM} A tungsten-halogen white-light source was used to illuminate the samples through an inverted microscope whose output was imaged on the entrance slit of a $F=320$mm spectrometer with a resolution of 0.5 nm. A 10$\times$ (0.25 N.A.) objective was used to collect the normally incident light for spectroscopic imaging onto a liquid nitrogen-cooled detector. A high-reflectance mirror (Newport Broadband SuperMirror, $R\geq99.9\%$) was used to normalize all signals. In addition, data from $\sim1000\mu$m across the films were averaged to obtain each final trace, in order to eliminate spurious effects from local inhomogeneities in the rough films. The spectral response of the film shown in Fig.\ref{SEMfig} is depicted in Fig.\ref{RTAfig}.

\section{Reflection, Transmission, and Absorption of random percolation composites}
Many metal-dielectric composite systems are described by conventional effective medium techniques (EMTs)\cite{MaxGarnett,Bruggman,MiltonBook,PodolskiyBook} by representing the composite as an effective homogeneous layer which successfully models the system's average optical properties.  However, it is known that the optical properties of these films close to the percolation threshold cannot be adequately described by EMTs.\cite{Yagil91,Yagil92}   The reason for the consistent failure of EMTs in this case is two-fold. First, although the dimensions of the components in percolation films are much smaller than the free-space wavelength, the optical properties of the composites are dominated by the dynamics of resonant clusters that can be comparable in size to the wavelength. Second, as result of a \textit{dc} metal-dielectric phase transition, the effective parameters of the percolation films in vicinity of $p_c$ become scale-dependent and therefore cannot be described by quasi-static effective medium models. Although some percolation films have been successfully described in terms of generalized Ohm's law (GOL)\cite{sarychevGOL,NathansohnGOL,BergmanGOL}, straightforward extensions of GOL formalism to our system are not consistent with our experimental observations.

The only technique that explicitly accounts for the \textit{dc} conductivity phase transition of percolation systems is known as {\it scaling theory}.\cite{Yagil91,Yagil92,Straley,anomdiff,ImadaRMP} In this technique, the conductivity of the film is assumed to be explicitly dependent on the size of the cluster $L$ over which it is measured. Specifically, the average conductivity of a conductive cluster of size $L$ is given by
\begin{widetext}
\begin{equation}
\label{SigmaM}
\sigma_m (L) = \frac{C_1 \sigma_{dc}}{1+ \omega^2 \tau^2} \left(\frac{L}{\xi_0}\right)^{-\mu/\nu}
+ i \left[\frac{C_1 \sigma_{dc} \omega \tau}{1+\omega^2 \tau^2} \left(\frac{L}{\xi_0}\right)^{-\mu/\nu} - C_2 \omega C_0 \left(\frac{L}{\xi_0}\right)^{s/\nu}\right],
\end{equation}
and that of a dielectric (insulating) cluster is
\begin{equation}
\label{SigmaD}
\sigma_d (L) = \frac{C_3 \omega^2 C_0^2}{\sigma_{dc}} \left(\frac{L}{\xi_0}\right)^{(\mu+2s)/\nu}
+ i \left[\frac{C_3 \omega^2 C_0^2 \omega \tau}{\sigma_{dc}} \left(\frac{L}{\xi_0}\right)^{(\mu+2s)/\nu} - C_4 \omega C_0 \left(\frac{L}{\xi_0}\right)^{s/\nu}\right].
\end{equation}
\end{widetext}

The expressions above explicitly assume that the conductivity of the conductive component of the film is given by the Drude model,
\begin{equation}
\label{drudesigma}
\sigma_{1}=\frac{\sigma_{dc}}{1-i \omega \tau},
\end{equation}
where $\sigma_{dc}$ is the dc conductivity, $\tau$ is the electron relaxation time, and $\omega$ is the angular frequency of the incident light. The $ac$ response of a dielectric film component is equivalent to that of a capacitor,
\begin{equation}
\label{capsigma}
\sigma_{2}= - i \omega C_0,
\end{equation}

\noindent where $C_0$ is the average capacitance between neighboring metal clusters.  The parameters $\sigma_{dc}, \tau, $ $\xi_0$, $C_0$ and $C_1\ldots C_4$ coefficients are uniquely determined by the composition and micro-geometry of the percolation film.  The critical exponents for $2$D percolating films are $\mu=s=1.3$ and $\nu=4/3$.\cite{Yagil91,Yagil92,MiriamPRL}  For $p \ll p_c$ percolation films are governed by dielectric conductivity, which is dominated by the capacitance coefficients $C_3$ and $C_4$.  For $p \gg p_c$ metallic conductivity (governed by $C_1$, $C_2$) dominates the optical properties of the system.

Despite the scale-dependence on the microscopic and mesoscopic levels, the percolation film appears homogeneous when measured over a significantly large area.  The transition from the scale-dependent to the homogeneous $dc$ response occurs at the scale known as the {\it correlation length}, $\xi$, that characterizes the typical cluster size. In the vicinity of the percolation threshold, the correlation length diverges as
\begin{equation}
\label{Xip}
\xi = \xi_0 \left|\frac{p-p_c}{p_c}\right|^{-\nu}.
\end{equation}

\noindent The constant $\xi_0$ represents the smallest metal cluster size, which occurs at $p\rightarrow 0$.

At finite frequencies, the oscillatory motion of electrons within conducting clusters leads to the length scale correction of the homogeneous response of the system,
\begin{equation}
\label{Lomega}
L(\lambda_0) = \min\left\{
\begin{array}{l}
B_0 \xi_0 \left(\lambda_0 / 2 \pi \xi_0 \right)^{1/(2+\theta)},
\\
\xi(p)
\end{array}
\right.
\end{equation}
where $\theta=0.79$, $B_0=4.0$, and the free space wavelength is given by $\lambda_0$.\cite{Yagil91,Yagil92}

The $ac$ conductivity of the percolation films, calculated using the expressions above can be directly related to an effective film index, which along with the film thickness $d$ can be used to determine the macroscopic optical properties of the film, including $R$ and $T$.
In our calculations, we use the technique introduced in Ref.[\onlinecite{Yagil91}]. In this approach, the optical properties of the film are calculated as a weighted average of conductive (dielectric) film contributions, where the average conductivities are given by Eqs.(\ref{SigmaM}) and (\ref{SigmaD}) respectively to yield

\begin{eqnarray}
\label{TransScaled}
T = \int_{0}^{\infty} \left[f T_\sigma(z \sigma_m)+ (1-f) T_\sigma (z \sigma_d)\right] P(z) dz
\\
R_i = \int_{0}^{\infty} [f R_{i,\sigma}(z \sigma_m) + (1-f) R_{i,\sigma} (z \sigma_d)] P(z) dz
\end{eqnarray}
where the parameter
\begin{equation}
\label{fSc}
f = \frac{1}{2}\left[1 + \left(\frac{p-p_c}{p_c}\right) \left(\frac{L}{\xi_0}\right)^{1/\nu} \right],
\end{equation}
is the metal occupation probability.  For small surface metal concentrations, $p<p_c\left[1-\left(\frac{L}{\xi_0}\right)^{-1/\nu}\right]$, the occupation probability $f \rightarrow 0$.  When $p>p_c\left[1+\left(\frac{L}{\xi_0}\right)^{-1/\nu}\right]$ the occupation probability $f \rightarrow 1$.  For intermediate surface metal concentrations centered at $p=p_c$ with full-width $\Delta p = 2p_c\left(\frac{L}{\xi_0}\right)^{-1/\nu}$, the occupation probability varies linearly as a function of p from unoccupied ($f=0$) to occupied ($f=1$).  As shown by the above inequalities, the range of metal surface coverage values for which scaled metal and dielectric optical properties are averaged depends non-trivially on the correlation length, applied frequency, and film geometry.  The function $P(z)$ gives the distribution of the conductivities of conductive [dielectric] clusters around their mean values given by Eq.(\ref{SigmaM}) [Eq.(\ref{SigmaD})]. Following Ref.[\onlinecite{Yagil91}] and [\onlinecite{epsexpansion}] we assume that $P(z)$ is adequately described by a log-normal distribution function with standard deviation of $\sigma_{sd} = 0.3$.  Integrating over all scaled conductivities averages out the length dependent optical conductivity and allows for percolation films to be modeled by the contributions from planar homogeneous constituent layers.

The homogeneous-layer optical properties are given by,\cite{EMfoundations}

\begin{eqnarray}
T_\sigma = \left|\frac{4 n_{\rm f} \sqrt{n_{\rm s}} \Phi}{(1+n_{\rm f})(n_{\rm f}+n_{\rm s})+(1-n_{\rm f})(n_{\rm f}-n_{\rm s})\Phi^2}\right|^2
\\
R_{1,\sigma} = \left|\frac{(1-n_{\rm f})(n_{\rm f}+n_{\rm s})+(n_{\rm f}-n_{\rm s})(1+n_{\rm f})\Phi^2}{(1+n_{\rm f})(n_{\rm f}+n_{\rm s})+(1-n_{\rm f})(n_{\rm f}-n_{\rm s})\Phi^2}\right|^2
\\
R_{2,\sigma} = \left|\frac{(n_{\rm f}-n_{\rm s})(1+n_{\rm f})+(1-n_{\rm f})(n_{\rm f}+n_{\rm s})\Phi^2}{(1+n_{\rm f})(n_{\rm f}+n_{\rm s})+(1-n_{\rm f})(n_{\rm f}-n_{\rm s})\Phi^2}\right|^2
\end{eqnarray}

\noindent where the glass substrate index is $n_{\rm s} = 1.5166$, the effective film index is $n_{\rm f}=\sqrt{1+4 \pi i \sigma/\omega}$, and the phase parameter is $\Phi = \exp (i \frac{\omega}{c} n_{\rm f} d)$.

As noted, all critical exponents in the expressions above are universal for all $2$D percolating networks, while the parameters $C_0\ldots C_4$, $\sigma_{dc}, \tau$,  and $\xi_0$ are unique for a given percolation film.\cite{Yagil91,Yagil92}  In our calculations, we use $\sigma_{dc}=2.574 \times 10^{17} $sec$^{-1}$, frequency dependent relaxation time\cite{Theye} $1 / \tau = 1/\tau_0 +\beta \omega^2$, where $\tau_0=3.0$fs, $\beta=0.2$fs, $C_0=0.5$, $C_1=C_2=0.046, C_3=0.028, C_4=0.055$, and $\xi_0=2$nm.

\section{comparison with experimental results and discussion}
\begin{figure}[h]
\includegraphics[width=\columnwidth]{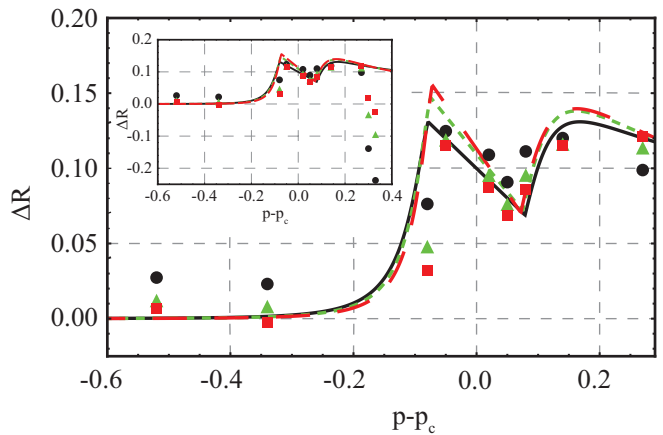}
\caption{(Color online) Points represent the measured change in reflectance ($\Delta R=R_1-R_2$) for various incident wavelengths. Black circles $500$nm, green triangles $600$nm, red boxes $700$nm.  Corresponding colored solid lines (black solid $500$nm, green short-dashed $600$nm, red long-dashed $700$nm) represent the results of scaling theory reflectance calculations.  The inset shows the change in reflectance over the entire surface coverage range.  Note that the 2D scaling model fails for large metal concentrations, where the three-dimensional structure of the composite dominates the optical response.
 \label{DeltaR} }
\end{figure}

\begin{figure}[h]
\includegraphics[width=\columnwidth]{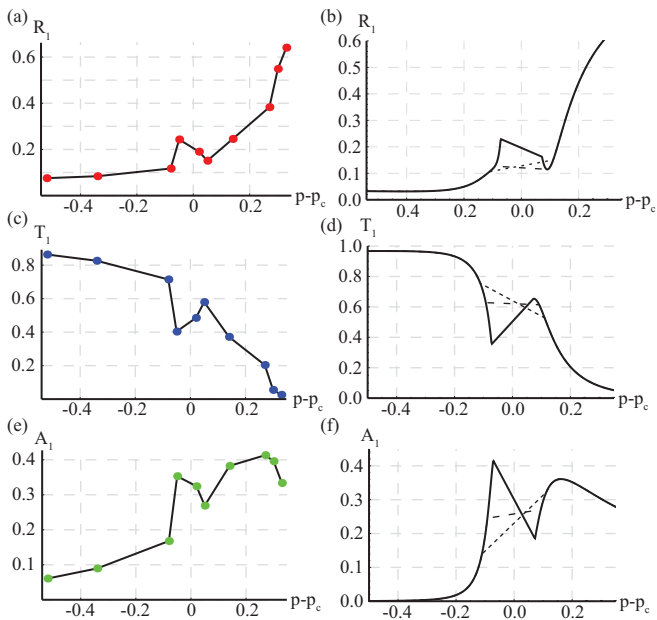}
\caption{(Color online) Points represent the measured (a) reflectance, (c) transmittance, and (e) loss from the air side as a function surface coverage fraction.  Connecting lines are a guide for the eye.  Calculated (b) reflectance, (d) transmittance, and (f) loss when the correlation length parameter is $\xi_0=2$nm (solid line), $\xi_0=5$nm (dashed line), and $\xi_0=10$nm (dotted line).  For all graphs the incident wavelength is $700$nm.
\label{RTpdata} }
\end{figure}

A comparison of the experimentally obtained spectral response of the silver films with the predictions of scaling theory is shown in Fig.\ref{RTAfig}.  It is seen that both the broadband nature of the reflectance asymmetry and its non-monotonic behavior near the percolation threshold are are well reproduced by the theoretical model, as demonstrated in Fig.\ref{DeltaR}.  We note however, that the model fails for the large metal concentrations $p\rightarrow 1$\cite{chenMM,JAPfilms}
where the structure of the composite becomes substantially three-dimensional and cannot be treated as a thin homogeneous
film.  To further illustrate the robustness of the presented technique we show in Fig.\ref{RTpdata} a comparison of experimentally measured values of $R_1$ and $T_1$, as well as the losses (computed as $A_1=1-R_1-T_1$,) with our theoretical model.  As mentioned before, both theoretical and experimental results clearly show that despite a strong reflectance asymmetry, the transmittance of the films remains symmetric.  Therefore the asymmetry in reflectance is directly related to the asymmetry in losses ($\Delta R = R_1 - R_2 = A_2 - A_1$).\cite{chenMM}

We now examine the loss (or, alternately the reflectance) in more detail.  We note that several previous experiments as well as theoretical models have observed absorption maxima in the vicinity of $p=p_c$. In contrast, our experimental data clearly demonstrate a local \emph{minimum} in losses near the percolation threshold, as seen in Fig.\ref{RTpdata}(e).  Additionally we observe a local \emph{maximum} in reflectance and a local \emph{minimum} in transmittance near $p=p_c$, as seen in Fig.\ref{RTpdata}(a,c).

We suggest that this anomalous behavior stems from the dramatically reduced correlation length in our solution-derived percolation films.  The smallest metallic particle size produced in our experiments is on the order of 2 nm, in contrast to 10 nm reported in Ref.[\onlinecite{Yagil92}]. In addition, the optical response of solution-derived metals is typically affected by the reduced electron mean free path\cite{electronMFP}, which may further reduce the effective particle size, described by the parameter $\xi_0$.

Figure \ref{RTpdata} demonstrates the evolution of optical properties of percolation systems when the correlation length is reduced, corresponding to a change in $\xi_0$ from 10 nm to 2 nm. It is clearly seen that at $\xi_0\simeq 2nm$ the absorption reaches a local minimum in the vicinity of $p=p_c$, while at $\xi_0=10$nm, the system recovers the absorption maximum at $p=p_c$, as observed in previous studies.

The dramatic difference between correlation lengths in our {\it solution-derived} films\cite{chenMM,JAPfilms} and other {\it vacuum-deposited} counterparts\cite{sarychevGOL,Yagil91,gadenne} is consistent with the difference in fabrication technique: while thermal deposition under vacuum typically yields uniform metal films with almost perfect Ohmic contacts between adjacent grains, solution-based deposition is routinely associated with quantitatively weaker contacts between conducting grains. The latter results in films with reduced electron mean-free-paths and lower correlation lengths.

\section{Conclusion}
We have developed an analytical description of the phenomenon  of broadband asymmetric reflection in percolation composites. The developed technique, based on scaling theory, is not only capable of describing the spectral response of our films, but also explains that the reduced correlation length in our solution-derived composites is the primary cause of the experimentally observed anomalous optical properties near the percolation threshold.  Our work demonstrates that the correlation length is an important factor that fundamentally affects the optical properties of percolation composites in the vicinity of the percolation threshold.

\section{Acknowledgements}
This research is sponsored in part by ONR (Grant No. N00014-07-1-0457), NSF (Grant No. ECCS-0724763), PRF (Grant No. 43924-G10), NSF DMR-02-39273, NSF DMR-08-04433.


\begin{thebibliography}{34}
\expandafter\ifx\csname natexlab\endcsname\relax\def\natexlab#1{#1}\fi
\expandafter\ifx\csname bibnamefont\endcsname\relax
  \def\bibnamefont#1{#1}\fi
\expandafter\ifx\csname bibfnamefont\endcsname\relax
  \def\bibfnamefont#1{#1}\fi
\expandafter\ifx\csname citenamefont\endcsname\relax
  \def\citenamefont#1{#1}\fi
\expandafter\ifx\csname url\endcsname\relax
  \def\url#1{\texttt{#1}}\fi
\expandafter\ifx\csname urlprefix\endcsname\relax\def\urlprefix{URL }\fi
\providecommand{\bibinfo}[2]{#2}
\providecommand{\eprint}[2][]{\url{#2}}

\bibitem[{\citenamefont{Stockman et~al.}(2001)\citenamefont{Stockman, Faleev,
  and Bergman}}]{StockmanLocal}
\bibinfo{author}{\bibfnamefont{M.~I.} \bibnamefont{Stockman}},
  \bibinfo{author}{\bibfnamefont{S.~V.} \bibnamefont{Faleev}},
  \bibnamefont{and} \bibinfo{author}{\bibfnamefont{D.~J.}
  \bibnamefont{Bergman}}, \bibinfo{journal}{Phys. Rev. Lett.}
  \textbf{\bibinfo{volume}{87}}, \bibinfo{pages}{167401}
  (\bibinfo{year}{2001}).

\bibitem[{\citenamefont{Genov et~al.}(2003)\citenamefont{Genov, Sarychev, and
  Shalaev}}]{genov}
\bibinfo{author}{\bibfnamefont{D.~A.} \bibnamefont{Genov}},
  \bibinfo{author}{\bibfnamefont{A.~K.} \bibnamefont{Sarychev}},
  \bibnamefont{and} \bibinfo{author}{\bibfnamefont{V.~M.}
  \bibnamefont{Shalaev}}, \bibinfo{journal}{Phys. Rev. E}
  \textbf{\bibinfo{volume}{67}}, \bibinfo{pages}{056611}
  (\bibinfo{year}{2003}).

\bibitem[{\citenamefont{Sarychev et~al.}(1999)\citenamefont{Sarychev, Shubin,
  and Shalaev}}]{akslocal}
\bibinfo{author}{\bibfnamefont{A.~K.} \bibnamefont{Sarychev}},
  \bibinfo{author}{\bibfnamefont{V.~A.} \bibnamefont{Shubin}},
  \bibnamefont{and} \bibinfo{author}{\bibfnamefont{V.~M.}
  \bibnamefont{Shalaev}}, \bibinfo{journal}{Phys. Rev. B}
  \textbf{\bibinfo{volume}{60}}, \bibinfo{pages}{16389} (\bibinfo{year}{1999}).

\bibitem[{\citenamefont{Sarychev and Shalaev}(2000)}]{physreports}
\bibinfo{author}{\bibfnamefont{A.~K.} \bibnamefont{Sarychev}} \bibnamefont{and}
  \bibinfo{author}{\bibfnamefont{V.~M.} \bibnamefont{Shalaev}},
  \bibinfo{journal}{Physics Reports} \textbf{\bibinfo{volume}{335}},
  \bibinfo{pages}{275} (\bibinfo{year}{2000}).

\bibitem[{\citenamefont{Shalaev and Sarychev}(1998)}]{shalaevNL}
\bibinfo{author}{\bibfnamefont{V.~M.} \bibnamefont{Shalaev}} \bibnamefont{and}
  \bibinfo{author}{\bibfnamefont{A.~K.} \bibnamefont{Sarychev}},
  \bibinfo{journal}{Phys. Rev. B} \textbf{\bibinfo{volume}{57}},
  \bibinfo{pages}{13265} (\bibinfo{year}{1998}).

\bibitem[{\citenamefont{Drachev et~al.}(2001)\citenamefont{Drachev, Bragg,
  Podolskiy, Safonov, Kim, Ying, Armstrong, and Shalaev}}]{optactiv}
\bibinfo{author}{\bibfnamefont{V.~P.} \bibnamefont{Drachev}},
  \bibinfo{author}{\bibfnamefont{W.~D.} \bibnamefont{Bragg}},
  \bibinfo{author}{\bibfnamefont{V.~A.} \bibnamefont{Podolskiy}},
  \bibinfo{author}{\bibfnamefont{V.~P.} \bibnamefont{Safonov}},
  \bibinfo{author}{\bibfnamefont{W.~T.} \bibnamefont{Kim}},
  \bibinfo{author}{\bibfnamefont{Z.~C.} \bibnamefont{Ying}},
  \bibinfo{author}{\bibfnamefont{R.~L.} \bibnamefont{Armstrong}},
  \bibnamefont{and} \bibinfo{author}{\bibfnamefont{V.~M.}
  \bibnamefont{Shalaev}}, \bibinfo{journal}{J. Opt. Soc. Am. B}
  \textbf{\bibinfo{volume}{18}}, \bibinfo{pages}{1896} (\bibinfo{year}{2001}).

\bibitem[{\citenamefont{Lagarkov et~al.}(1997)\citenamefont{Lagarkov, Rozanov,
  Sarychev, and Simonov}}]{lagarkov}
\bibinfo{author}{\bibfnamefont{A.}~\bibnamefont{Lagarkov}},
  \bibinfo{author}{\bibfnamefont{K.}~\bibnamefont{Rozanov}},
  \bibinfo{author}{\bibfnamefont{A.}~\bibnamefont{Sarychev}}, \bibnamefont{and}
  \bibinfo{author}{\bibfnamefont{N.}~\bibnamefont{Simonov}},
  \bibinfo{journal}{Physica A} \textbf{\bibinfo{volume}{241}},
  \bibinfo{pages}{199} (\bibinfo{year}{1997}).

\bibitem[{\citenamefont{Shalaev}(2002)}]{shalaevbook}
\bibinfo{author}{\bibfnamefont{V.~M.} \bibnamefont{Shalaev}},
  \emph{\bibinfo{title}{Optical Properties of Nanostructured Random Media}}
  (\bibinfo{publisher}{Springer}, \bibinfo{year}{2002}), \bibinfo{edition}{1st}
  ed.

\bibitem[{\citenamefont{Stockman}(2000)}]{Stockmanprl2000}
\bibinfo{author}{\bibfnamefont{M.~I.} \bibnamefont{Stockman}},
  \bibinfo{journal}{Phys. Rev. Lett.} \textbf{\bibinfo{volume}{84}},
  \bibinfo{pages}{1011} (\bibinfo{year}{2000}).

\bibitem[{\citenamefont{Stockman et~al.}(1994)\citenamefont{Stockman, Pandey,
  Muratov, and George}}]{Stockmanprl1994}
\bibinfo{author}{\bibfnamefont{M.~I.} \bibnamefont{Stockman}},
  \bibinfo{author}{\bibfnamefont{L.~N.} \bibnamefont{Pandey}},
  \bibinfo{author}{\bibfnamefont{L.~S.} \bibnamefont{Muratov}},
  \bibnamefont{and} \bibinfo{author}{\bibfnamefont{T.~F.}
  \bibnamefont{George}}, \bibinfo{journal}{Phys. Rev. Lett.}
  \textbf{\bibinfo{volume}{72}}, \bibinfo{pages}{2486} (\bibinfo{year}{1994}).

\bibitem[{\citenamefont{Chen et~al.}(2007)\citenamefont{Chen, Hasegawa,
  Podolskiy, and Deutsch}}]{chenMM}
\bibinfo{author}{\bibfnamefont{A.}~\bibnamefont{Chen}},
  \bibinfo{author}{\bibfnamefont{K.}~\bibnamefont{Hasegawa}},
  \bibinfo{author}{\bibfnamefont{V.~A.} \bibnamefont{Podolskiy}},
  \bibnamefont{and} \bibinfo{author}{\bibfnamefont{M.}~\bibnamefont{Deutsch}},
  \bibinfo{journal}{Opt. Lett.} \textbf{\bibinfo{volume}{32}},
  \bibinfo{pages}{1770} (\bibinfo{year}{2007}).

\bibitem[{\citenamefont{Stauffer and Aharony}(1992)}]{intropt}
\bibinfo{author}{\bibfnamefont{D.}~\bibnamefont{Stauffer}} \bibnamefont{and}
  \bibinfo{author}{\bibfnamefont{A.}~\bibnamefont{Aharony}},
  \emph{\bibinfo{title}{Introduction to Percolation Theory}}
  (\bibinfo{publisher}{Taylor {\&} Francis, London}, \bibinfo{year}{1992}),
  \bibinfo{edition}{2nd} ed.

\bibitem[{\citenamefont{Barron}(2004)}]{ChiralBook}
\bibinfo{author}{\bibfnamefont{L.~D.} \bibnamefont{Barron}},
  \emph{\bibinfo{title}{Molecular light scattering and optical activity}}
  (\bibinfo{publisher}{Cambridge University Press, Cambridge},
  \bibinfo{year}{2004}), \bibinfo{edition}{2nd} ed.

\bibitem[{\citenamefont{Fedotov et~al.}(2006)\citenamefont{Fedotov, Mladyonov,
  Prosvirnin, Rogacheva, Chen, and Zheludev}}]{chiralprl}
\bibinfo{author}{\bibfnamefont{V.~A.} \bibnamefont{Fedotov}},
  \bibinfo{author}{\bibfnamefont{P.~L.} \bibnamefont{Mladyonov}},
  \bibinfo{author}{\bibfnamefont{S.~L.} \bibnamefont{Prosvirnin}},
  \bibinfo{author}{\bibfnamefont{A.~V.} \bibnamefont{Rogacheva}},
  \bibinfo{author}{\bibfnamefont{Y.}~\bibnamefont{Chen}}, \bibnamefont{and}
  \bibinfo{author}{\bibfnamefont{N.~I.} \bibnamefont{Zheludev}},
  \bibinfo{journal}{Phys. Rev. Lett.} \textbf{\bibinfo{volume}{97}},
  \bibinfo{pages}{167401} (\bibinfo{year}{2006}).

\bibitem[{foo()}]{footnoteLoss}
\bibinfo{note}{Since the model we use here utilizes a uniform smooth film of
  known thickness, standard boundary conditions allow only specular reflection
  to occur. We therefore employ the common approach which does not distinguish
  between specular and diffuse loss mechanisms,\cite{Yagil91} lumping them
  together into a \emph{generalized combined loss}.}

\bibitem[{\citenamefont{{M. S. M. Peterson} and Deutsch}(2009)}]{JAPfilms}
\bibinfo{author}{\bibnamefont{{M. S. M. Peterson}}} \bibnamefont{and}
  \bibinfo{author}{\bibfnamefont{M.}~\bibnamefont{Deutsch}},
  \bibinfo{journal}{J. Appl. Phys.} \textbf{\bibinfo{volume}{106}},
  \bibinfo{pages}{063722} (\bibinfo{year}{2009}).

\bibitem[{\citenamefont{{J. C. Maxwell Garnett}}(1904)}]{MaxGarnett}
\bibinfo{author}{\bibnamefont{{J. C. Maxwell Garnett}}},
  \bibinfo{journal}{Philos. Trans. R. Soc. London A}
  \textbf{\bibinfo{volume}{203}}, \bibinfo{pages}{385} (\bibinfo{year}{1904}).

\bibitem[{\citenamefont{Bruggeman}(1935)}]{Bruggman}
\bibinfo{author}{\bibfnamefont{D.}~\bibnamefont{Bruggeman}},
  \bibinfo{journal}{Ann. Phys. (Leipzig)} \textbf{\bibinfo{volume}{24}},
  \bibinfo{pages}{636} (\bibinfo{year}{1935}).

\bibitem[{\citenamefont{Milton}(2002)}]{MiltonBook}
\bibinfo{author}{\bibfnamefont{G.~W.} \bibnamefont{Milton}},
  \emph{\bibinfo{title}{The Theory of Composites}}
  (\bibinfo{publisher}{Cambridge University Press}, \bibinfo{year}{2002}),
  \bibinfo{edition}{1st} ed.

\bibitem[{\citenamefont{Noginov and Podolskiy}(2011)}]{PodolskiyBook}
\bibinfo{author}{\bibfnamefont{M.~A.} \bibnamefont{Noginov}} \bibnamefont{and}
  \bibinfo{author}{\bibfnamefont{V.~A.} \bibnamefont{Podolskiy}},
  \emph{\bibinfo{title}{Tutorials in Metamaterials}} (\bibinfo{publisher}{CRC
  Press}, \bibinfo{year}{2011}), \bibinfo{edition}{1st} ed.

\bibitem[{\citenamefont{Yagil et~al.}(1991)\citenamefont{Yagil, Yosefin,
  Bergman, Deutscher, and Gadenne}}]{Yagil91}
\bibinfo{author}{\bibfnamefont{Y.}~\bibnamefont{Yagil}},
  \bibinfo{author}{\bibfnamefont{M.}~\bibnamefont{Yosefin}},
  \bibinfo{author}{\bibfnamefont{D.~J.} \bibnamefont{Bergman}},
  \bibinfo{author}{\bibfnamefont{G.}~\bibnamefont{Deutscher}},
  \bibnamefont{and} \bibinfo{author}{\bibfnamefont{P.}~\bibnamefont{Gadenne}},
  \bibinfo{journal}{Phys. Rev. B} \textbf{\bibinfo{volume}{43}},
  \bibinfo{pages}{11342} (\bibinfo{year}{1991}).

\bibitem[{\citenamefont{Yagil et~al.}(1992)\citenamefont{Yagil, Gadenne,
  Julien, and Deutscher}}]{Yagil92}
\bibinfo{author}{\bibfnamefont{Y.}~\bibnamefont{Yagil}},
  \bibinfo{author}{\bibfnamefont{P.}~\bibnamefont{Gadenne}},
  \bibinfo{author}{\bibfnamefont{C.}~\bibnamefont{Julien}}, \bibnamefont{and}
  \bibinfo{author}{\bibfnamefont{G.}~\bibnamefont{Deutscher}},
  \bibinfo{journal}{Phys. Rev. B} \textbf{\bibinfo{volume}{46}},
  \bibinfo{pages}{2503} (\bibinfo{year}{1992}).

\bibitem[{\citenamefont{Sarychev et~al.}(1995)\citenamefont{Sarychev, Bergman,
  and Yagil}}]{sarychevGOL}
\bibinfo{author}{\bibfnamefont{A.~K.} \bibnamefont{Sarychev}},
  \bibinfo{author}{\bibfnamefont{D.~J.} \bibnamefont{Bergman}},
  \bibnamefont{and} \bibinfo{author}{\bibfnamefont{Y.}~\bibnamefont{Yagil}},
  \bibinfo{journal}{Phys. Rev. B} \textbf{\bibinfo{volume}{51}},
  \bibinfo{pages}{5366} (\bibinfo{year}{1995}).

\bibitem[{\citenamefont{Levy-Nathansohn and
  Bergman}(1997{\natexlab{a}})}]{NathansohnGOL}
\bibinfo{author}{\bibfnamefont{R.}~\bibnamefont{Levy-Nathansohn}}
  \bibnamefont{and} \bibinfo{author}{\bibfnamefont{D.~J.}
  \bibnamefont{Bergman}}, \bibinfo{journal}{Physica A}
  \textbf{\bibinfo{volume}{241}}, \bibinfo{pages}{166}
  (\bibinfo{year}{1997}{\natexlab{a}}).

\bibitem[{\citenamefont{Levy-Nathansohn and
  Bergman}(1997{\natexlab{b}})}]{BergmanGOL}
\bibinfo{author}{\bibfnamefont{R.}~\bibnamefont{Levy-Nathansohn}}
  \bibnamefont{and} \bibinfo{author}{\bibfnamefont{D.~J.}
  \bibnamefont{Bergman}}, \bibinfo{journal}{Phys. Rev. B}
  \textbf{\bibinfo{volume}{55}}, \bibinfo{pages}{5425}
  (\bibinfo{year}{1997}{\natexlab{b}}).

\bibitem[{\citenamefont{Straley}(1976)}]{Straley}
\bibinfo{author}{\bibfnamefont{J.~P.} \bibnamefont{Straley}},
  \bibinfo{journal}{J. Phys. C} \textbf{\bibinfo{volume}{9}},
  \bibinfo{pages}{783} (\bibinfo{year}{1976}).

\bibitem[{\citenamefont{Gefen et~al.}(1983)\citenamefont{Gefen, Aharony, and
  Alexander}}]{anomdiff}
\bibinfo{author}{\bibfnamefont{Y.}~\bibnamefont{Gefen}},
  \bibinfo{author}{\bibfnamefont{A.}~\bibnamefont{Aharony}}, \bibnamefont{and}
  \bibinfo{author}{\bibfnamefont{S.}~\bibnamefont{Alexander}},
  \bibinfo{journal}{Phys. Rev. Lett.} \textbf{\bibinfo{volume}{50}},
  \bibinfo{pages}{77} (\bibinfo{year}{1983}).

\bibitem[{\citenamefont{Imada et~al.}(1998)\citenamefont{Imada, Fujimori, and
  Tokura}}]{ImadaRMP}
\bibinfo{author}{\bibfnamefont{M.}~\bibnamefont{Imada}},
  \bibinfo{author}{\bibfnamefont{A.}~\bibnamefont{Fujimori}}, \bibnamefont{and}
  \bibinfo{author}{\bibfnamefont{Y.}~\bibnamefont{Tokura}},
  \bibinfo{journal}{Rev. Mod. Phys.} \textbf{\bibinfo{volume}{70}},
  \bibinfo{pages}{1039} (\bibinfo{year}{1998}).

\bibitem[{\citenamefont{Rohde et~al.}(2006)\citenamefont{Rohde, Hasegawa, and
  Deutsch}}]{MiriamPRL}
\bibinfo{author}{\bibfnamefont{C.~A.} \bibnamefont{Rohde}},
  \bibinfo{author}{\bibfnamefont{K.}~\bibnamefont{Hasegawa}}, \bibnamefont{and}
  \bibinfo{author}{\bibfnamefont{M.}~\bibnamefont{Deutsch}},
  \bibinfo{journal}{Phys. Rev. Lett.} \textbf{\bibinfo{volume}{96}},
  \bibinfo{pages}{045503} (\bibinfo{year}{2006}).

\bibitem[{\citenamefont{Rammal et~al.}(1985)\citenamefont{Rammal, Lemieux, and
  {A. M. S. Tremblay}}}]{epsexpansion}
\bibinfo{author}{\bibfnamefont{R.}~\bibnamefont{Rammal}},
  \bibinfo{author}{\bibfnamefont{M.~A.} \bibnamefont{Lemieux}},
  \bibnamefont{and} \bibinfo{author}{\bibnamefont{{A. M. S. Tremblay}}},
  \bibinfo{journal}{Phys. Rev. Lett.} \textbf{\bibinfo{volume}{54}},
  \bibinfo{pages}{1087} (\bibinfo{year}{1985}).

\bibitem[{\citenamefont{Reitz et~al.}(1993)\citenamefont{Reitz, Milford, and
  Christy}}]{EMfoundations}
\bibinfo{author}{\bibfnamefont{J.~R.} \bibnamefont{Reitz}},
  \bibinfo{author}{\bibfnamefont{F.~J.} \bibnamefont{Milford}},
  \bibnamefont{and} \bibinfo{author}{\bibfnamefont{R.~W.}
  \bibnamefont{Christy}}, \emph{\bibinfo{title}{Foundations of Electromagnetic
  Theory}} (\bibinfo{publisher}{Addison-Wesley}, \bibinfo{year}{1993}),
  \bibinfo{edition}{4th} ed.

\bibitem[{\citenamefont{Theye}(1970)}]{Theye}
\bibinfo{author}{\bibfnamefont{M.~L.} \bibnamefont{Theye}},
  \bibinfo{journal}{Phys. Rev. B} \textbf{\bibinfo{volume}{2}},
  \bibinfo{pages}{3060} (\bibinfo{year}{1970}).

\bibitem[{\citenamefont{Lisseberger and Nelson}(1974)}]{electronMFP}
\bibinfo{author}{\bibfnamefont{P.~H.} \bibnamefont{Lisseberger}}
  \bibnamefont{and} \bibinfo{author}{\bibfnamefont{R.~G.}
  \bibnamefont{Nelson}}, \bibinfo{journal}{Thin Solid Films}
  \textbf{\bibinfo{volume}{21}}, \bibinfo{pages}{159} (\bibinfo{year}{1974}).

\bibitem[{\citenamefont{Gadenne et~al.}(1989)\citenamefont{Gadenne, Yagil, and
  Deutscher}}]{gadenne}
\bibinfo{author}{\bibfnamefont{P.}~\bibnamefont{Gadenne}},
  \bibinfo{author}{\bibfnamefont{Y.}~\bibnamefont{Yagil}}, \bibnamefont{and}
  \bibinfo{author}{\bibfnamefont{G.}~\bibnamefont{Deutscher}},
  \bibinfo{journal}{J. Appl. Phys.} \textbf{\bibinfo{volume}{66}},
  \bibinfo{pages}{3019} (\bibinfo{year}{1989}).

\end{thebibliography}

\end{document}